# A Broadband Algorithm for Adiabatic Mode Evolution and its Application on Polarization Splitter-Rotator on LNOI Platform

Geng Chen, Chijun Li, Xuanhao Wang, An Pan, Junjie Wei, Yuankang Huang, Siyu Lu, Yiqi Dai, Xiangyu Meng, Cheng Zeng and Jinsong Xia

## Abstract

Adiabatic mode evolution waveguides (AMEWs) are widely utilized in integrated photonics, including tapered waveguides, edge couplers, mode converters, splitters, etc. An analytical theory and a novel AMEW design algorithm are developed to create shortcuts to adiabaticity (STA). This new algorithm is effective in shortening the total length of the AMEW while maintaining the desired wavelength range. Moreover, this analytical algorithm requires much fewer computing resources than traditional numerical algorithms. With the new algorithm, we demonstrate a broadband and highly efficient polarization splitter-rotator (PSR) on a lithium-niobate-on-insulator (LNOI) platform with an LN thickness of 500 nm. According to our simulation, the length of the PSR is shortened by 3.5 times compared to the linear design. The fabricated PSR, with a total length of 2 mm, exhibits an insertion loss (IL) of 0.8 dB and a polarization extinction ratio (ER) of 12.2 dB over a wavelength range exceeding 76 nm.

## Introduction

Adiabatic mode evolution waveguides (AMEWs) convert power between the guiding modes along the waveguide without mode leakage, if the adiabatic length is sufficient and the propagation constant varies slowly. AMEWs are highly versatile in integrated photonics, offering wide bandwidth (BW) and high robustness, and are used in various applications such as curves [1–3], tapers [4–6], edge couplers [7–9], mode splitters and rotators [10–12], etc. However, most AMEWs are designed with linear tapers, and the lengths of the devices are often very large. Linear and segmented-linear AMEWs provide only a few degrees of freedom in the design process [13], severely restricting their potential for parametric optimization. While existing linear designs are adequate for many applications on the silicon-on-insulator (SOI) platform due to strong light confinement, AMEWs on lower refractive index difference platforms like lithium-niobate-on-insulator (LNOI) sometimes tend to be excessively long [14] for photonic integrated circuits (PICs). For example, polarization splitter-rotators (PSRs) on 500-nm-thick LNOI platform [14] are considered more challenging compared to 300-nm [15] or 400-nm thickness [16], as the refractive index difference $\Delta n_{\text{eff}}$ at the mode hybridization point is significantly smaller. According to our analysis, there is an inverse square relationship between AMEW length and refractive index difference. $\Delta n_{\text{eff}}$ is very sensitive to the thickness ($\Delta n_{\text{eff}}$ at the hybridization point is about 0.01 on 400-nm-thick LNOI [16] and only about 0.0022 in this work as illustrated in Fig. **3**(b)). As observed in various phenomena, e.g. Rabi splitting [17,18] and dual microring resonators coupling [19,20], the difference between the eigenvalues depends on the coupling strength of the two states. Thinner films are much more conducive to generating strong coupling supermodes for two reasons: (1) Due to the sidewall angle of LNOI waveguides, thinner films allow waveguides to be closer to each other; (2) The mode hybridization point is located in smaller $w_1$ (0.4 μm on 400-nm-thick LNOI [16] while 0.76 μm on 500-nm-thick LNOI at 1.55μm wavelength), leading to larger mode overlap. In [14], the total length of the AMEWs on a 500-nm-thick LNOI platform is 7 mm, much longer than that on a 400-nm-thick platform [16]. Consequently, effective shortcuts to adiabaticity (STA) design algorithms are essential for AMEWs on such low refractive index platforms.

Typically, fully numerical algorithms have been used to design STA for AMEWs [21,22], guided by Eigenmode Expansion (EME) simulation results. However, such approaches are time-consuming and demand substantial computing resources. Additionally, the fully numerical algorithms are not impractical for most devices on LNOI due to the overlong lengths (ranging from hundreds of microns to several millimeters), which consume excessive computing memory. Moreover, many numerical algorithms focus on a specific wavelength [21,23] while BW is increasingly critical in modern PICs. Although there are analytical algorithms for the STA design [24–26], they often rely on some assumptions to simplify waveguide structures, limiting their applicability. The fast quasi-adiabatic (FAQUAD) algorithm [27], an analytical algorithm, has been applied to SOI AMEWs [4,28]. This algorithm is free of simplification of the waveguide structure and requires only Finite Difference Eigenmode (FDE) simulation for the STA design. However, this algorithm was initially designed for adiabatic quantum computing, so the reasonability of applying it to AMEWs lacks detailed theoretical analysis. Besides, previous works also focus on a single wavelength, and a broadband design strategy is also highly needed. To address this issue, we first propose the adiabatic waveguide approximation (AWA), which offers a detailed analysis of the mode evolution inside a slowly varying waveguide. Then based on the AWA and the classical FAQUAD, we propose an innovative design algorithm referred to as the broadband FAQUAD to achieve a broad and flexible wavelength range design.

This paper proposes a novel broadband FAQUAD algorithm, providing a BW-flexible design strategy and saving computing resources. To demonstrate the effect of the broadband FAQUAD, we design and fabricate the linear and STA PSRs on 500-nm-thick x-cut LNOI. Simulation results show that the length of the STA PSR is reduced by a factor of 3.5 compared to the linear design. Both simulation and experimental results indicate that the STA PSR exhibits superior performance and a shorter length. The STA PSR achieves an extinct ratio (ER) of 12.2 dB and an insertion loss (IL) of 0.8 dB over a wide wavelength range from 1.554 μm to beyond 1.63 μm.

## The Adiabatic Waveguide Approximation

The adiabatic quantum approximation (AQA), first demonstrated by M. Born and V. Fock in 1928 [29], is commonly used in adiabatic quantum computing [30,31] and annealing [32]. The AQA states that *a system remains in its instantaneous eigenstate if a given perturbation is acting on it slowly enough*. Although the physics of the Schrödinger and the Helmholtz equations are entirely different, their mathematical nature is similar, because they are both eigenequations (as demonstrated in Table 1). This similarity inspires us to apply the AQA to optics and develop the AWA, starting with the Helmholtz equation without further simplification or assumption on the waveguide structure. The AWA states a similar conclusion to the quantum-mechanical situation: *the electromagnetic wave remains in its instantaneous eigenmode if the perturbation of the waveguide is small enough, i.e. the AMEW length is large enough* (see SI).

Detailed theoretical analysis indicates that the total length of AMEWs should be long enough to realize the adiabaticity (see SI)

$$L \gg C \cdot A_{mn} \tag{1}$$

Where $C$ is a slowly varying term so we consider it as a constant and:

$$A_{mn} = \frac{I_{mn}}{\Delta n_{\mathrm{eff},mn}^2} \frac{dw_1}{d\zeta} \tag{2}$$

With $I_{mn}, \Delta n_{\mathrm{eff},mn}^2$ being the line integration (see SI) and the effective index difference between the two modes $m$ and $n$, and $\zeta = z/L$ is the normalized propagation length.

**Table 1** Correspondence with Quantum Mechanics

| | Quantum Mechanics | Waveguide Theory |
|---|---|---|
| Basic Equation | $i\hbar \frac{d}{dt}\lvert\Psi\rangle = \hat{H}\lvert\Psi\rangle$ | $\nabla^2\Psi + k^2\Psi = 0$ |
| Longitudinal Direction | $t$ | $z$ |
| Transverse Direction | $x, y, z$ | $x, y$ |
| Eigen equation | $\hat{H}\lvert n\rangle = E_n\lvert n\rangle$ | $-\left(\nabla_t^2 + k^2\right)\psi_n = -\beta_n^2\psi_n$ |
| Eigenstate | $\lvert n\rangle$ | $\psi_n(x, y)$ |
| Eigenvalue | $E_n$ | $-\beta_n^2$ |

## III. Basic Structure and the Design Algorithm

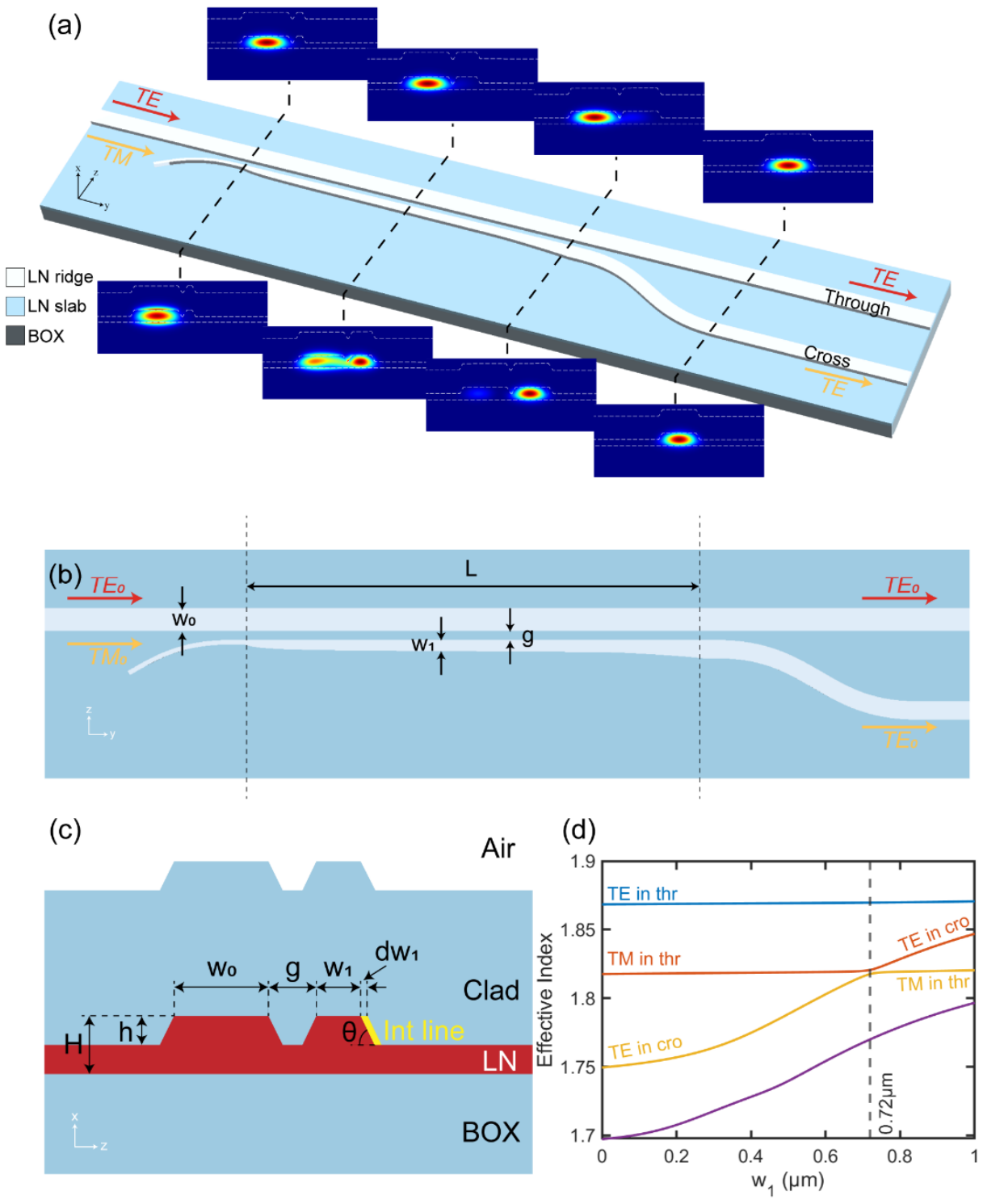

**Fig. 1.** (a) Schematic of the PSR. The top four insets are the adiabatic mode evolution process of TE input. The mode remains in the through waveguide along the propagation. The bottom four insets are the adiabatic mode evolution process of TE input. The mode is transferred into the cross waveguide during propagation. (b) Vertical view of the PSR, with $w_0 = 1.3$ μm, $g = 0.45$ μm, and $w_1$ varying from 0.3 μm to 1 μm. (c) Waveguide cross-section, with $H = 0.5$ μm, $h = 0.26$ μm and $\theta = 60°$. Buried oxide (BOX) and clad are both $SiO_2$. The integration line is indicated as the yellow line on the right boundary of the cross waveguide (see SI). (d) The evolution of the first four modes along the widening $w_1$. $n_{\text{eff}}$ of TM in through waveguide is larger than TE in drop at first but surpassed when $w_1 > 0.72$ μm, while $n_{\text{eff}}$ of TE remains the largest in the entire evolution.

The next paragraph will introduce a detailed description of the broadband FAQUAD algorithm, after a demonstration of the basic structure of the PSR. The schematic (Fig. 1(a)) and vertical view (Fig. 1(b)) illustrate the basic structure of the PSR [16]. The waveguides propagate along the y-axis of the LN crystal. The top width of the through waveguide $w_0$ is constantly 1.3 μm, while the cross waveguide top width $w_1$ widens gradually from 0.3 μm to 1 μm (top width). The top gap $g$ between the two waveguides is 0.45 μm. The waveguide cross-section is depicted in Fig. 1(c), showing a ridge waveguide height ($H$) of 500 nm and an etch depth ($h$) of 260 nm. The sidewall angle ($\theta$) is 60°, consistent with our fabrication process. The quasi-transverse-electric mode (TE) in the through waveguide has the largest $n_{\text{eff}}$ during the widening of $w_1$ so the energy stays in the through waveguide. Conversely, the quasi-transverse-magnetic mode (TM) in the through waveguide gradually transfers to TE in the cross waveguide during the process that the $n_{\text{eff}}$ of TE in the cross waveguide surpasses that of TM in the through waveguide, as shown in Fig. 1(d). The mode evolution process is fully illustrated in the insets of Fig. 1(a).

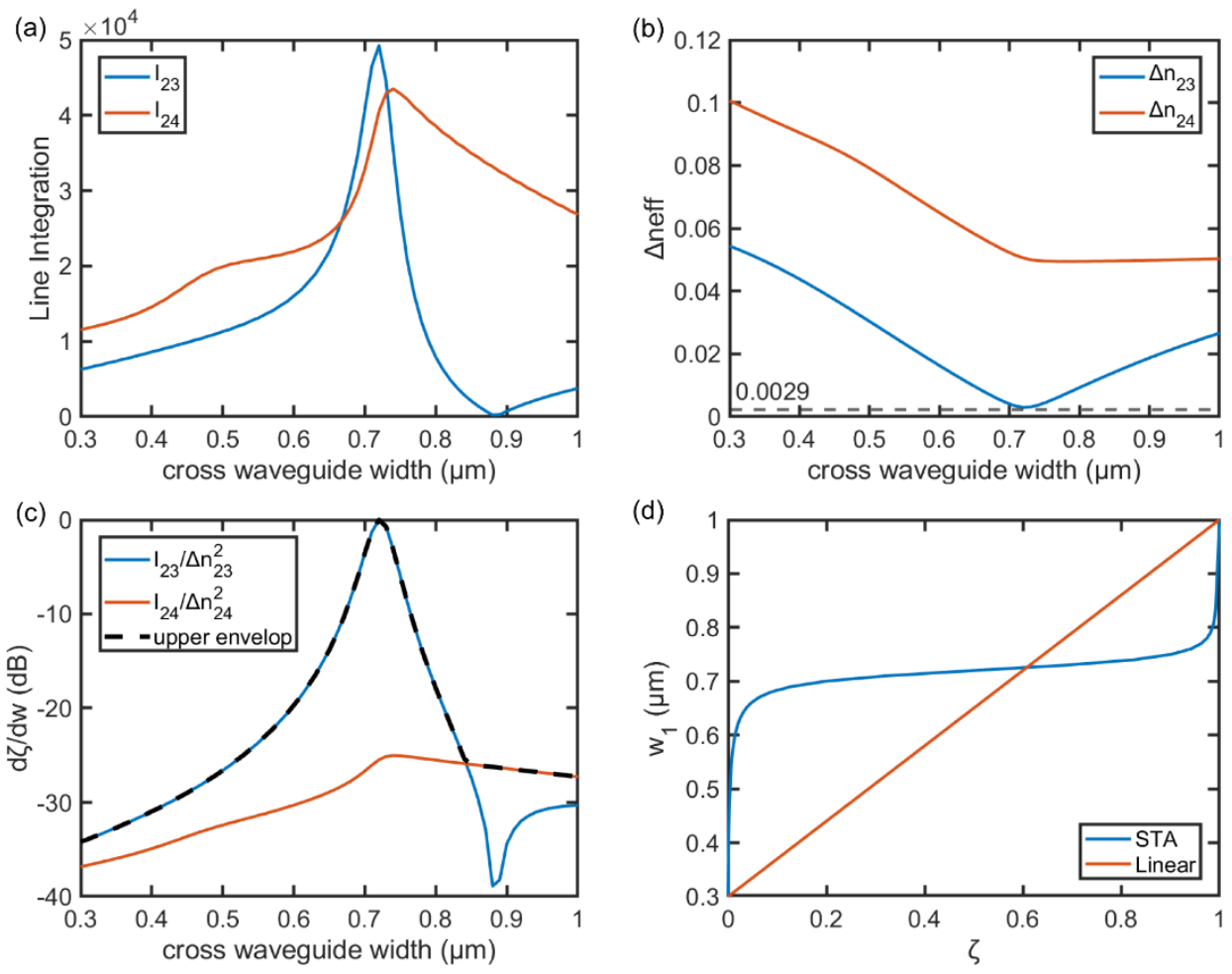

**Fig. 2.** PSR design parameters at a single wavelength of 1.6 μm. (a) the line integration as a function of $w_1$. (b) the refractive index difference as a function of $w_1$. The dashed line indicates a minimal value of 0.0029. (c) Calculated $I_{23}/\Delta n_{\text{eff},23}^2$, $I_{24}/\Delta n_{\text{eff},24}^2$ and their upper envelop as functions of $w_1$. (d) The line shape of STA compared to linear design.

Based on the basic PSR structure and (2), if the AMEW varies linearly (i.e. $dw_1/d\zeta$ is constant), $A_{mn}$ forms a "hill" in the middle of the propagation, as illustrated in Fig. 2(c). To flatten $A_{mn}$ and shorten the AMEW, we introduce the concept of FAQUAD [27]:

$$\frac{d\zeta}{dw_1} \propto \frac{I_{mn}}{\Delta n_{\text{eff},mn}^2} \tag{3}$$

So that $A_{mn}$ is invariant during propagation. To ensure the adiabaticity of TM input, we consider $m = 2, n = 3,4$ in the PSR instance and perform a simulation in FDE. Fig. 2(a) (b) illustrate simulated $I_{23}, I_{24}$ and $\Delta n_{\text{eff},23}, \Delta n_{\text{eff},24}$ as functions of $w_1$ at a wavelength of 1.6 μm, then (c) (d) are the calculated $I_{23}/\Delta n_{\text{eff},23}^2$, $I_{24}/\Delta n_{\text{eff},24}^2$ and the STA curve shape. The FAQUAD method only ensures effective conversion in a single wavelength. To design a broadband PSR, a simple way is to select a set of wavelengths in the target range and take their upper envelope, allowing for a flexible design strategy for BW:

$$\frac{d\zeta}{dw_1} = \max\left\{\frac{I_{2n}^{\lambda_j}}{\left(\Delta n_{\text{eff},2n}^{\lambda_j}\right)^2}\right\} \tag{4}$$

A set of wavelengths from 1.55 to 1.65 μm is used to calculate the broadband PSR, as depicted in Fig. 3. This final curve shape is used for fabrication and characterization, and the device length is 2 mm.

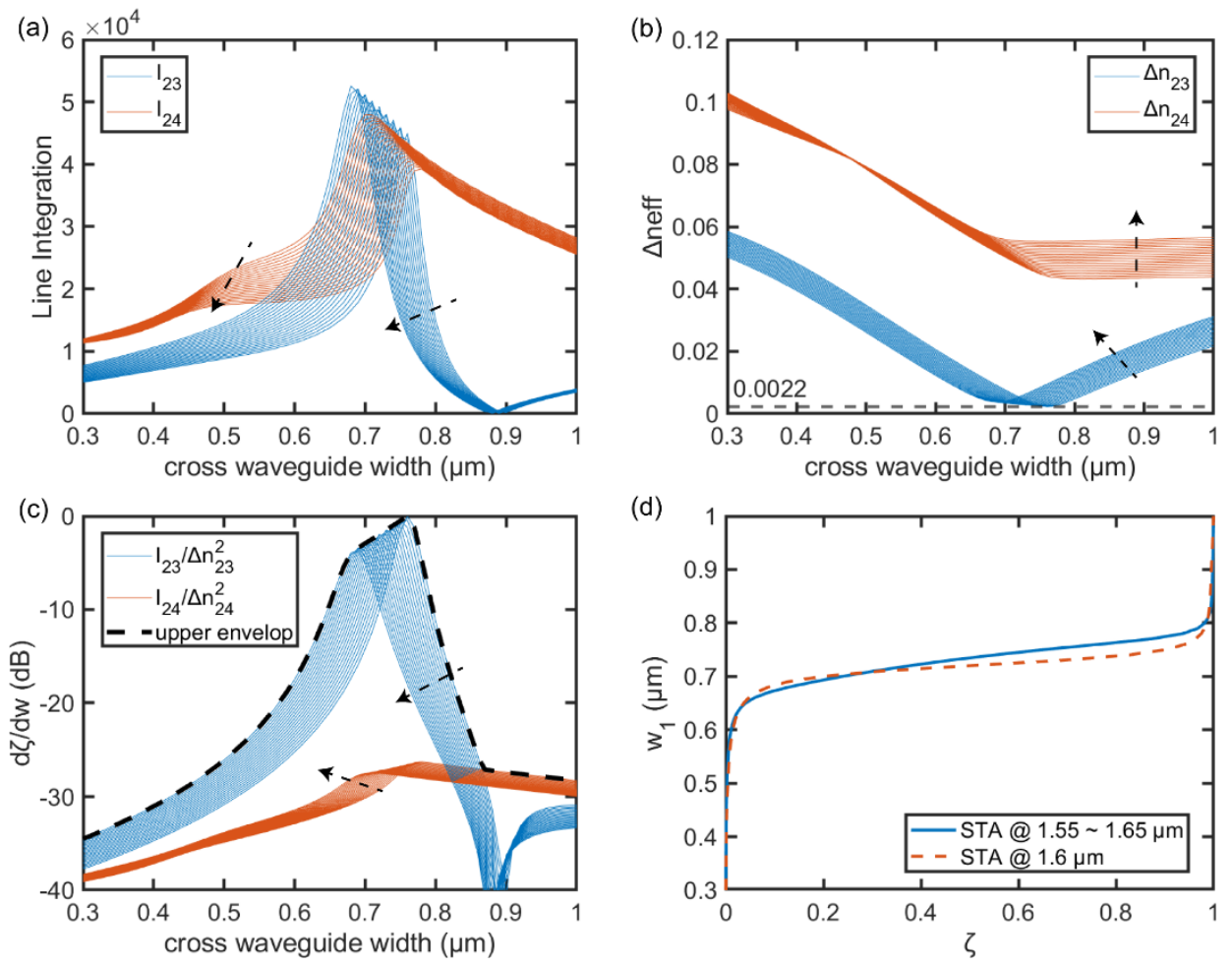

**Fig. 3.** PSR design parameters at a broad wavelength range of 1.55 ~ 1.65 μm. (a) the line integration as a function of $w_1$. (b) the refractive index difference as a function of $w_1$. The dashed line indicates a minimal value of 0.0022. (c) Calculated $I_{23}/\Delta n_{\text{eff},23}^2$, $I_{24}/\Delta n_{\text{eff},24}^2$ and their upper envelope as functions of $w_1$. (d) The line shape of broadband STA compared to single wavelength STA design. All arrows in (a), (b), and (c) indicate the wavelength change from 1.55 to 1.65 μm.

The linear- and STA-design simulation results of field evolution (Fig. 4(a)), the transmission of TM input to cross port (Fig. 4(b)), and the wavelength sweep of the transmission of TM input to through port (Fig. 4(c)) are presented. According to the simulation, the broadband FAQUAD algorithm significantly shortens the PSR by 3.5 times at the 1.6 μm wavelength and has a better performance over the target wavelength range compared to the linear design.

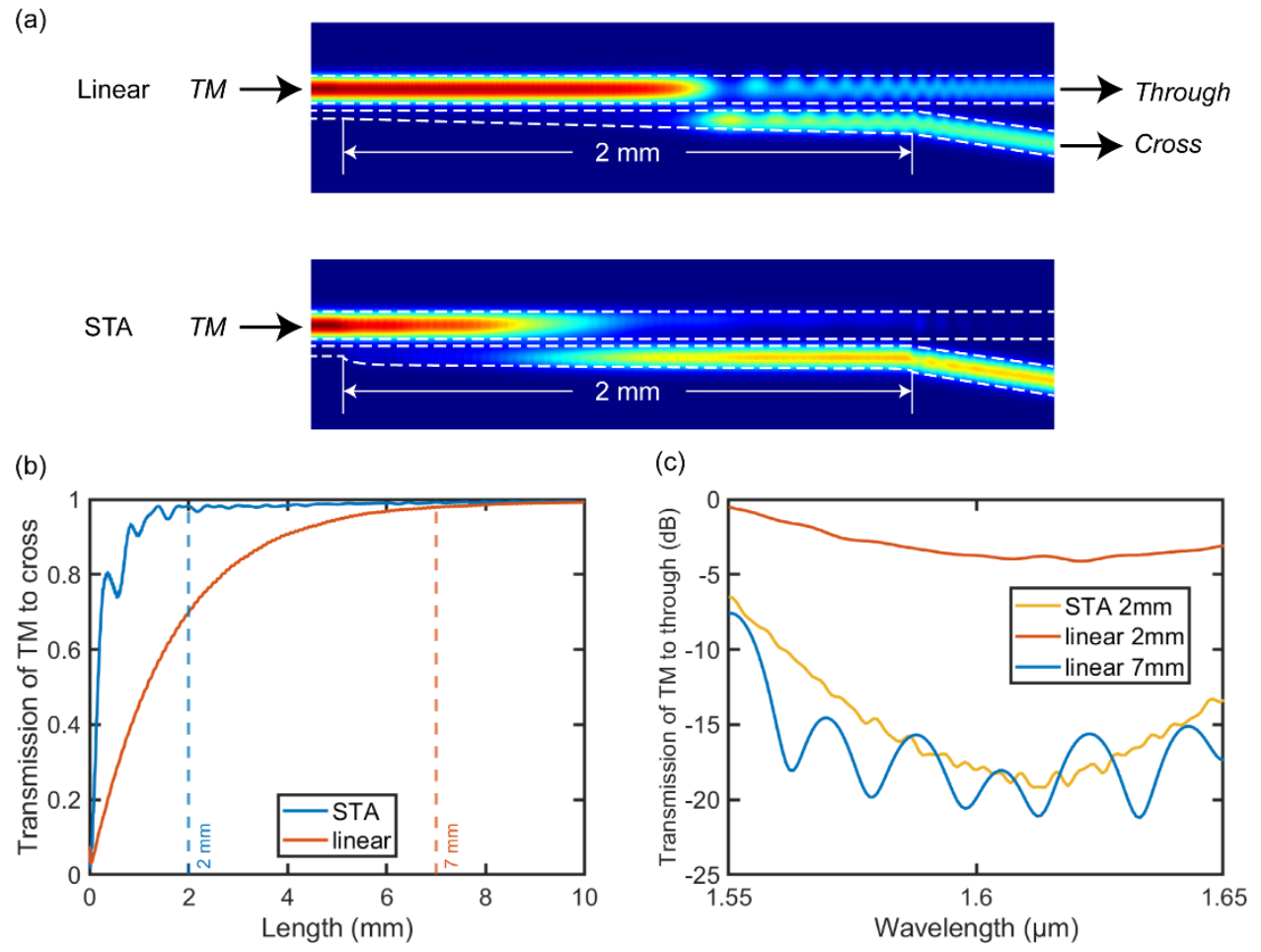

**Fig. 4.** Simulation Results. (a) Field evolution of linear and STA design at 1.6 μm. The STA design shows a higher conversion efficiency at the same length. (b) Conversion efficiency

of STA and linear design when TM input (1.6 μm). Vertical lines indicate 98% conversion efficiency to cross port. (c) BW simulation results of TM input to through port with STA design at 2 mm length and linear design at 2mm, 7mm length.

# IV. Device Fabrication and Characterization

To characterize the performance of the broadband FAQUAD, both linear and broadband-STA PSRs are fabricated on a 500-nm-thick x-cut LNOI wafer with a 4.7-μm buried oxide layer (NanoLN), with lengths of 2 mm. The patterns of PSRs on the substrate were achieved using electron beam lithography (EBL) and inductively coupled plasma (ICP) etching, followed by the deposition of a 1-μm-thick $SiO_2$ layer on the LN via plasma-enhanced chemical vapor deposition (PECVD). Two identical PSRs with different grating couplers are fabricated for TE and TM input, respectively. The microscopy and SEM images of the fabricated PSRs are shown in Fig. 5.

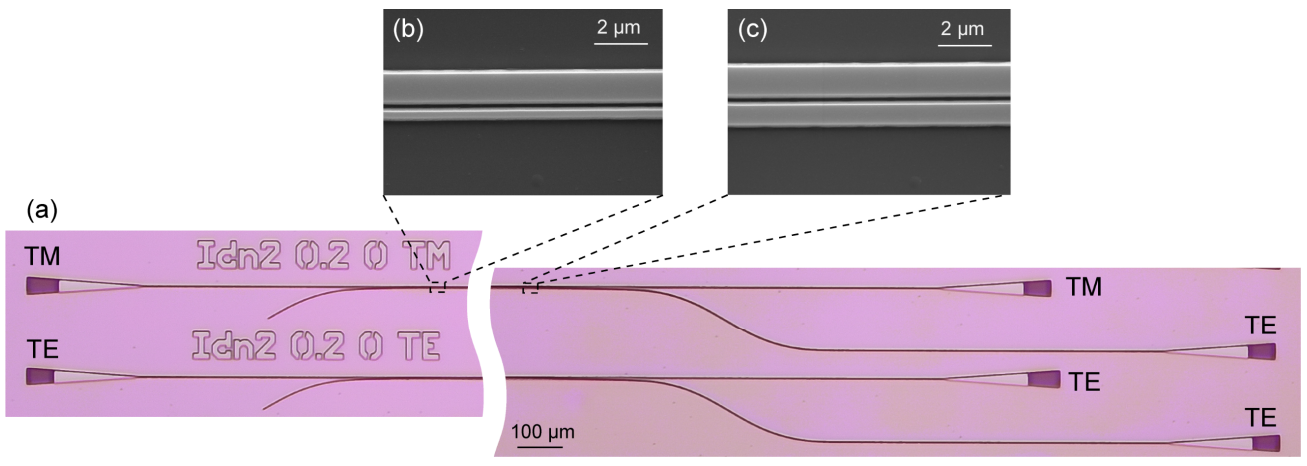

**Fig. 5.** (a) Microscopy image of the STA design PSRs. The GC types for different input polarization are labeled. (b) SEM image of the start of the AMEW. (c) SEM image of the end of the AMEW.

As depicted in Fig. 6(a), a tunable laser (TL, Santec TSL-550) is used to generate a continuous-wave (CW) pump ranging from 1.5 to 1.63 μm, then a fiber polarization controller (FPC) is connected to adjust the input polarization. The light coupled out from the chip is collected by a power meter (PM, Santec MPM-210). The experimental results for both linear and STA design are shown in Fig. 6, in which (b), (c) are for the linear PSR and (d), (e) are for the STA design PSR. The insertion loss of the devices is < 0.8 dB within the target wavelength range except for the TM input of linear design, which is due to the low conversion efficiency. The ERs of the TE input are 10.2 dB for the linear PSR and 12.2 dB for the STA design PSR, showing similar performances. However, the linear PSR exhibits an ER of less than 3.4 dB for TM input, corresponding to a conversion efficiency of only ~ 67%. The STA design achieved a large ER of 12.4 dB for TE and TM inputs over a BW exceeding 76 nm, matching the design target and the simulation results well. The test results support the efficiency of the broadband FAQUAD algorithm in shortening AMEWs and enhancing their performances. Besides, we also characterize linear 3-mm and 4-mm linear PSRs, their ERs of TM input are compared in Fig. 6(f). The linear PSRs have a low ER of less than 7.5 dB in the target wavelength range. Moreover, the ER exhibits no significant improvement as the length increases, suggesting that the linear PSRs are far from adiabaticity. Additionally, The STA design exhibits a much steeper BW cutoff of around 1.55 μm compared to the linear design, indicating that the shortening of the AMEW comes from sacrificing the device performance in the unwanted wavelength range.

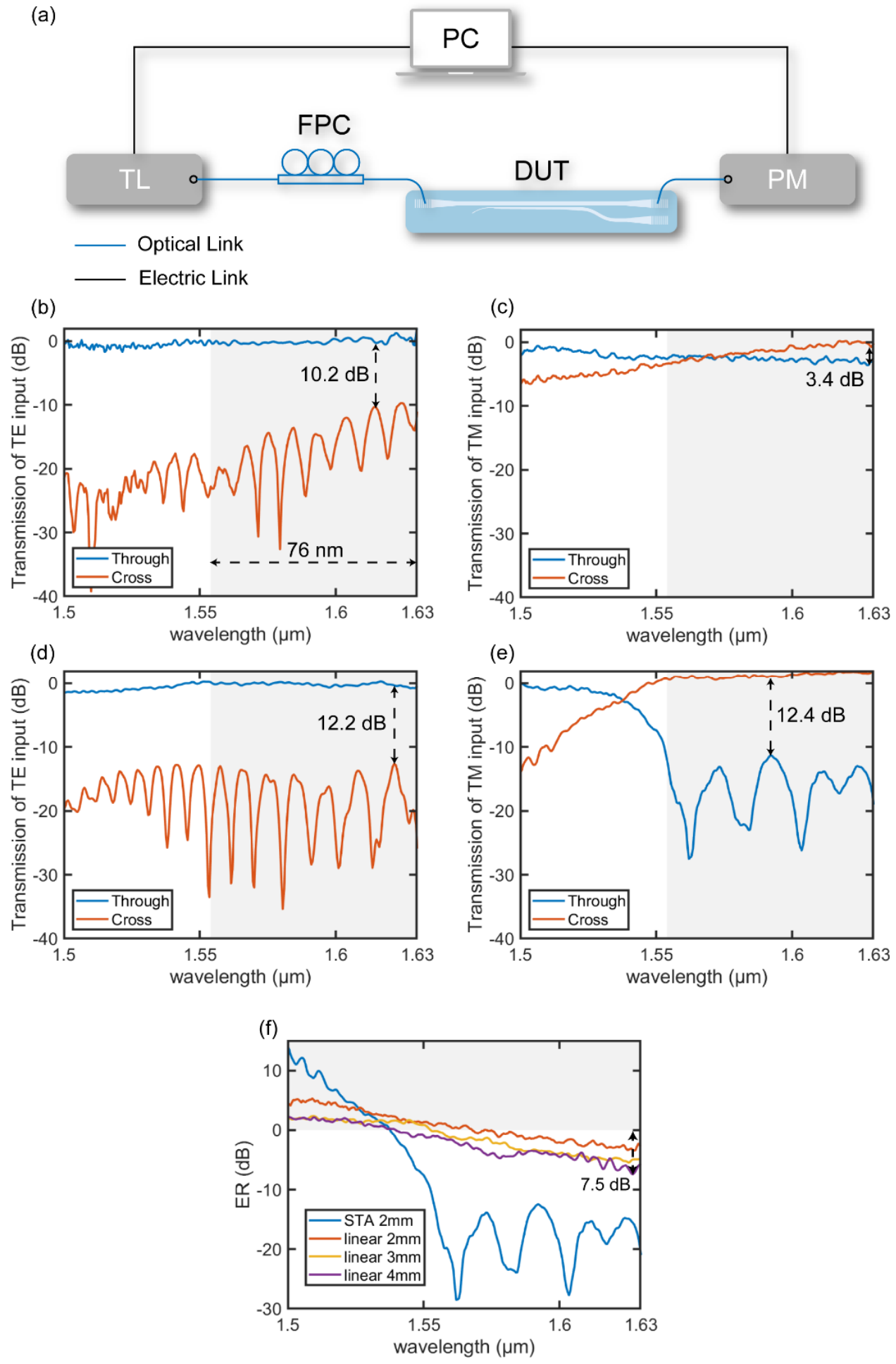

**Fig. 6.** (a) Test configuration. TL: Tunable Laser. FPC: Fiber Polarization Controller. DUT: Device Under Test. PM: Power Meter. PC: Personal Computer. And test results of (b) TE and (c) TM input of linear PSR. (d) TE and (e) TM input of STA PSR. The results show a 12 dB ER with an over 76-nm BW (1.554 ~ 1.63 μm corresponds to the shaded area). (f) ER comparison of STA and linear PSR when TM input. The ER is calculated by $T_{thr}(\text{dB}) - T_{cro}(\text{dB})$. The positive values in the shaded area are because the conversion efficiency is less than 50%.

**Table 2** Performances of several PSRs on LNOI

| Ref. | Shape | LN thickness/nm | L/ mm | ER/dB | IL/ dB | BW/nm |
|---|---|---|---|---|---|---|
| [15] | Segmented-linear & MMI | 360 | 0.6 | 10 | 1.5 | 47 |
| [16] | Linear | 400 | 1 | 20 | 0.5 | 110 |
| [14] | Segmented-linear | 500 | 7 | 10 | 3 | 130 |
| [33] | Segmented-linear & Y-junction | 500 | 0.44 | 19.6 | 1 | 60 |
| [34] | Segmented-linear & ADC | 600 | 0.24 | 10.6 | 0.91 | 50 |
| [35] | MMI | 600 | 1.6 | 9.5 | 3 | 39 |
| This work | STA | 500 | 2 | 12.2 | 0.8 | 76 |

Table 2 illustrates the comparison of several PSRs. Some PSRs exhibit large ERs, BWs, and short lengths because they are fabricated on a thinner LNOI platform [16] or propagate along the z-axis [34], having a larger $\Delta n_{\mathrm{eff}}$ at the mode hybridization point than our PSR thus being easier for PSR applications. The MMI [15,35], Y-junction [33], and ADC [34] approaches provide shorter device lengths but limited BW. Our STA PSR shows a balanced length, ER, BW, and IL performance.

# V. Discussion and Conclusion

This work builds a bridge between the AWA and the AQA [36]. Developed for decades, more accurate theoretical approximations, estimations [37], and advanced algorithms [38] used in AQA are also available for designing STA AMEWs, promising more strategies to further enhance their performances.

In conclusion, a novel design strategy is developed for designing AMEWs, providing significant length reduction and flexible wavelength range selection. We then fabricate and test a PSR using this strategy, and both simulation and test results have demonstrated the effectiveness of the broadband FAQUAD in reducing the total length of AMEWs while maintaining the desired wavelength range. A total BW of over 76 nm with a 12 dB ER is achieved with a length of only 2 mm.

# Reference

1. T. Chen, H. Lee, J. Li, and K. J. Vahala, "A general design algorithm for low optical loss adiabatic connections in waveguides," Opt. Express **20**, 22819–22829 (2012).
2. M. Cherchi, S. Ylinen, M. Harjanne, M. Kapulainen, and T. Aalto, "Dramatic size reduction of waveguide bends on a micron-scale silicon photonic platform," Opt. Express **21**, 17814–17823 (2013).
3. X. Jiang, H. Wu, and D. Dai, "Low-loss and low-crosstalk multimode waveguide bend on silicon," Opt. Express **26**, 17680–17689 (2018).
4. H.-C. Chung and S.-Y. Tseng, "Ultrashort and broadband silicon polarization splitter-rotator using fast quasiadiabatic dynamics," Opt. Express **26**, 9655–9665 (2018).
5. Y. Liu, W. Sun, H. Xie, N. Zhang, K. Xu, Y. Yao, S. Xiao, and Q. Song, "Adiabatic and Ultracompact Waveguide Tapers Based on Digital Metamaterials," IEEE J. Sel. Top. Quantum Electron. **25**, 1–6 (2019).
6. M. R. Karim, N. Al Kayed, R. Rafi, and B. M. A. Rahman, "Design and analysis of

inverse tapered silicon nitride waveguide for flat and highly coherent supercontinuum generation in the mid-infrared," Opt. Quantum Electron. **56**, 68 (2023).
7. I. Krasnokutska, J.-L. J. Tambasco, and A. Peruzzo, "Nanostructuring of LNOI for efficient edge coupling," Opt. Express **27**, 16578–16585 (2019).
8. A. He, X. Guo, K. Wang, Y. Zhang, and Y. Su, "Low Loss, Large Bandwidth Fiber-Chip Edge Couplers Based on Silicon-on-Insulator Platform," J. Light. Technol. **38**, 4780–4786 (2020).
9. B. Bhandari, C.-S. Im, K.-P. Lee, S.-M. Kim, M.-C. Oh, and S.-S. Lee, "Compact and Broadband Edge Coupler Based on Multi-Stage Silicon Nitride Tapers," IEEE Photonics J. **12**, 1–11 (2020).
10. W. D. Sacher, T. Barwicz, B. J. F. Taylor, and J. K. S. Poon, "Polarization rotator-splitters in standard active silicon photonics platforms," Opt. Express **22**, 3777–3786 (2014).
11. J. Wang, B. Niu, Z. Sheng, A. Wu, W. Li, X. Wang, S. Zou, M. Qi, and F. Gan, "Novel ultra-broadband polarization splitter-rotator based on mode-evolution tapers and a mode-sorting asymmetric Y-junction," Opt. Express **22**, 13565–13571 (2014).
12. H.-P. Chung, C.-H. Lee, K.-H. Huang, S.-L. Yang, K. Wang, A. S. Solntsev, A. A. Sukhorukov, F. Setzpfandt, and Y.-H. Chen, "Broadband on-chip polarization mode splitters in lithium niobate integrated adiabatic couplers," Opt. Express **27**, 1632–1645 (2019).
13. H. Luo, Z. Chen, H. Li, L. Chen, Y. Han, Z. Lin, S. Yu, and X. Cai, "High-Performance Polarization Splitter-Rotator Based on Lithium Niobate-on-Insulator Platform," IEEE Photonics Technol. Lett. **33**, 1423–1426 (2021).
14. Z. Chen, J. Yang, W.-H. Wong, E. Y.-B. Pun, and C. Wang, "Broadband adiabatic polarization rotator-splitter based on a lithium niobate on insulator platform," Photonics Res. **9**, 2319–2324 (2021).
15. C. Deng, W. Zhu, Y. Sun, M. Lu, L. Huang, D. Wang, G. Hu, B. Yun, and Y. Cui, "Broadband Polarization Splitter-Rotator on Lithium Niobate-on-Insulator Platform," IEEE Photonics Technol. Lett. **35**, 7–10 (2023).
16. R. Gan, L. Qi, Z. Ruan, J. Liu, C. Guo, K. Chen, and L. Liu, "Fabrication tolerant and broadband polarization splitter-rotator based on adiabatic mode evolution on thin-film lithium niobate," Opt. Lett. **47**, 5200–5203 (2022).
17. T. Yoshie, A. Scherer, J. Hendrickson, G. Khitrova, H. M. Gibbs, G. Rupper, C. Ell, O. B. Shchekin, and D. G. Deppe, "Vacuum Rabi splitting with a single quantum dot in a photonic crystal nanocavity," Nature **432**, 200–203 (2004).
18. H. Toida, T. Nakajima, and S. Komiyama, "Vacuum Rabi Splitting in a Semiconductor Circuit QED System," Phys. Rev. Lett. **110**, 066802 (2013).
19. M. Zhang, C. Wang, Y. Hu, A. Shams-Ansari, T. Ren, S. Fan, and M. Lončar, "Electronically programmable photonic molecule," Nat. Photonics **13**, 36–40 (2019).
20. A. Tikan, J. Riemensberger, K. Komagata, S. Hönl, M. Churaev, C. Skehan, H. Guo, R. N. Wang, J. Liu, P. Seidler, and T. J. Kippenberg, "Emergent nonlinear phenomena in a driven dissipative photonic dimer," Nat. Phys. **17**, 604–610 (2021).
21. T.-L. Liang, Y. Tu, X. Chen, Y. Huang, Q. Bai, Y. Zhao, J. Zhang, Y. Yuan, J. Li, F. Yi, W. Shao, and S.-T. Ho, "A Fully Numerical Method for Designing Efficient Adiabatic Mode Evolution Structures (Adiabatic Taper, Coupler, Splitter, Mode Converter) Applicable to Complex Geometries," J. Light. Technol. **39**, 5531–5547 (2021).

22. T.-L. Liang, X. Cheng, J. Shi, G. Wu, K. Xu, W. Rong, L. Lin, and W. Shao, "Analysis and Design of Compact Adiabatic Mode Converters Based on Adiabatic Mode Evolutions," J. Light. Technol. **41**, 6356–6361 (2023).
23. T. Liang, X. Cheng, M. Yu, L. Zhang, J. Shi, and W. Shao, "Numerical method for designing ultrashort and efficient adiabatic mode converters," JOSA B **39**, 2637–2642 (2022).
24. A. Milton and W. Burns, "Mode coupling in optical waveguide horns," IEEE J. Quantum Electron. **13**, 828–835 (1977).
25. T. A. Ramadan, R. Scarmozzino, and R. M. Osgood, "Adiabatic couplers: design rules and optimization," J. Light. Technol. **16**, 277–283 (1998).
26. X. Sun, H.-C. Liu, and A. Yariv, "Adiabaticity criterion and the shortest adiabatic mode transformer in a coupled-waveguide system," Opt. Lett. **34**, 280–282 (2009).
27. S. Martínez-Garaot, A. Ruschhaupt, J. Gillet, Th. Busch, and J. G. Muga, "Fast quasiadiabatic dynamics," Phys. Rev. A **92**, 043406 (2015).
28. C.-H. Chen, Y.-F. Lo, G.-X. Lu, Y.-J. Hung, and S.-Y. Tseng, "Adiabaticity Engineered Silicon Coupler With Design-Intended Splitting Ratio," IEEE Photonics J. **16**, 1–5 (2024).
29. M. Born and V. Fock, "Beweis des Adiabatensatzes," Z. Für Phys. **51**, 165–180 (1928).
30. R. Barends, A. Shabani, L. Lamata, J. Kelly, A. Mezzacapo, U. L. Heras, R. Babbush, A. G. Fowler, B. Campbell, Y. Chen, Z. Chen, B. Chiaro, A. Dunsworth, E. Jeffrey, E. Lucero, A. Megrant, J. Y. Mutus, M. Neeley, C. Neill, P. J. J. O'Malley, C. Quintana, P. Roushan, D. Sank, A. Vainsencher, J. Wenner, T. C. White, E. Solano, H. Neven, and J. M. Martinis, "Digitized adiabatic quantum computing with a superconducting circuit," Nature **534**, 222–226 (2016).
31. N. N. Hegade, K. Paul, Y. Ding, M. Sanz, F. Albarrán-Arriagada, E. Solano, and X. Chen, "Shortcuts to Adiabaticity in Digitized Adiabatic Quantum Computing," Phys. Rev. Appl. **15**, 024038 (2021).
32. S. Mandrà, Z. Zhu, and H. G. Katzgraber, "Exponentially Biased Ground-State Sampling of Quantum Annealing Machines with Transverse-Field Driving Hamiltonians," Phys. Rev. Lett. **118**, 070502 (2017).
33. X. Wang, A. Pan, T. Li, C. Zeng, and J. Xia, "Efficient polarization splitter-rotator on thin-film lithium niobate," Opt. Express **29**, 38044–38052 (2021).
34. Y. Wu, X. Sun, X. Xue, H. Li, S. Liu, Y. Zheng, and X. Chen, "Compact Adiabatic Polarization Splitter-Rotator on Thin-Film Lithium Niobate," J. Light. Technol. **42**, 2429–2435 (2024).
35. M. Wang, H. Yao, J. Deng, Z. Hu, T. Tang, and K. Chen, "Polarization splitter rotator on thin film lithium niobate based on multimode interference," (2024).
36. M. H. S. Amin, "Consistency of the Adiabatic Theorem," Phys. Rev. Lett. **102**, 220401 (2009).
37. L. C. Venuti, T. Albash, D. A. Lidar, and P. Zanardi, "Adiabaticity in open quantum systems," Phys. Rev. A **93**, 032118 (2016).
38. T. Albash and D. A. Lidar, "Adiabatic quantum computation," Rev. Mod. Phys. **90**, 015002 (2018).
39. M. V. Berry, "Quantal phase factors accompanying adiabatic changes," Proc. R. Soc. Lond. Math. Phys. Sci. **392**, 45–57 (1997).

## Supplementary Information

In waveguides, we consider the Helmholtz equation as the start:

$$\nabla^2\Psi(x,y,z)+k^2\Psi(x,y,z)=0 \tag{5}$$

Where $\nabla$ is the nabla operator, $\Psi$ is the electric or magnetic field, and $k = 2\pi n/\lambda$ is the wavenumber. To solve eigenmodes of waveguide in a certain propagation distance $z$, divide $\Psi$ as $\Psi(x,y,z) = \psi(x,y)Z(z)$, then:

$$\left(\nabla_t^2+k^2\right)\psi=\beta^2\psi \tag{6}$$

Where $\beta$ is the propagation constant and $\nabla_t^2 = \nabla^2 - d^2/dz^2$ is the transverse nabla operator. Rewriting (5), (6) in the form of Hamiltonian and Dirac notation:

$$\frac{d^2}{dz^2}\left|\Psi(z)\right\rangle=\hat{H}\left|\Psi(z)\right\rangle \tag{7}$$

$$\hat{H}(z)\left|n(z)\right\rangle=-\beta_n^2\left|n(z)\right\rangle \tag{8}$$

Where $\hat{H}(z) = -(\nabla_t^2 + k^2)$, $|\Psi(z)\rangle = \Psi(x,y,z)$, $|n(z)\rangle$ and $\beta_n(z)$ represent the n-th eigenmode and propagation constant in the distance $z$. Normalize (7), (8) with $\zeta = z/L$, where $L$ is the total length of the structure:

$$\frac{d^2}{d\zeta^2}\left|\tilde{\Psi}(\zeta)\right\rangle=L^2\tilde{H}(\zeta)\left|\tilde{\Psi}(\zeta)\right\rangle \tag{9}$$

$$\tilde{H}(\zeta)\left|\tilde{n}(\zeta)\right\rangle=-\tilde{\beta}_n^2(\zeta)\left|\tilde{n}(\zeta)\right\rangle \tag{10}$$

The tilde will be omitted in the following analysis for simplicity. Define the inner product as:

$$\langle\psi\mid\varphi\rangle=\iint_S\left(\mathbf{E}_\psi\times\mathbf{H}_\varphi^*+\mathbf{E}_\psi^*\times\mathbf{H}_\varphi\right)d\mathbf{s} \tag{11}$$

In which $\mathbf{E},\mathbf{H}$ are normalized field components then the orthogonality of different modes $\langle m|n\rangle = \delta_{mn}$ is satisfied. Here we introduce two crucial lemmas [29], [39]:

$$\langle m|\frac{d}{d\zeta}|n\rangle=\frac{1}{\beta_m^2-\beta_n^2}\langle m|\frac{dH}{d\zeta}|n\rangle \tag{12}$$

$$\mathrm{Re}\left\{\langle m|\frac{d}{d\zeta}|m\rangle\right\}=0 \tag{13}$$

Assume that the start state is:

$$\left|\Psi(0)\right\rangle=\sum_n c_n(0)e^{-i\theta_n(0)}\left|n(0)\right\rangle \tag{14}$$

Where $c_n(\zeta)$ is the mode expansion coefficient of the n-th eigenmode at $\zeta$ and $|c_n(\zeta)|^2$ represents the energy in the mode; $\theta_n$ is the phase cumulated during the propagation of the n-th mode and $d\theta_n/dz$, so:

$$\frac{d}{d\zeta}\theta_n(\zeta)=L\beta_n(\zeta) \tag{15}$$

Substituting (9) into (14):

$$\begin{aligned}\text{LHS} &= \sum_n \frac{d^2}{d\zeta^2}\left(c_n e^{-i\theta_n}\left|n\right\rangle\right) \\ &= \sum_n \left(\frac{d^2 c_n}{d\zeta^2} e^{-i\theta_n}\left|n\right\rangle - L^2\beta_n^2 c_n e^{-i\theta_n}\left|n\right\rangle \right. \\ &\quad + c_n e^{-i\theta_n}\frac{d^2\left|n\right\rangle}{d\zeta^2} - 2iL\beta_n \frac{dc_n}{d\zeta} e^{-i\theta_n}\left|n\right\rangle \\ &\quad \left. + 2\frac{dc_n}{d\zeta} e^{-i\theta_n}\frac{d\left|n\right\rangle}{d\zeta} - 2iL\beta_n c_n e^{-i\theta_n}\frac{d\left|n\right\rangle}{d\zeta}\right)\end{aligned} \tag{16}$$

$$\begin{aligned}\text{RHS} &= \sum_n H\left(c_n e^{-i\theta_n}\left|n\right\rangle\right) = \sum_n c_n e^{-i\theta_n} H\left|n\right\rangle \\ &= \sum_n -L^2\beta_n^2 c_n e^{-i\theta_n}\left|n\right\rangle\end{aligned} \tag{17}$$

Note that RHS equals the second term of LHS, and take the inner product with $|m\rangle$:

$$\begin{aligned} -\frac{1}{L}\frac{d^2 c_m}{d\zeta^2} + 2i\beta_m \frac{dc_m}{d\zeta} - 2\left(\frac{1}{L}\frac{dc_m}{d\zeta} - i\beta_m c_m\right)\left\langle m\right|\frac{d}{d\zeta}\left|m\right\rangle \\ = \sum_{n\neq m}\left\{ e^{i(\theta_m-\theta_n)}\left[\left(\frac{2}{L}\frac{dc_n}{d\zeta} - i\beta_n c_n\right)\left\langle m\right|\frac{d}{d\zeta}\left|n\right\rangle \right.\right. \\ \left.\left. + \frac{c_n}{L}\left\langle m\right|\frac{d^2}{d\zeta^2}\left|n\right\rangle\right]\right\}\end{aligned} \tag{18}$$

Here we apply two approximations. Firstly, as $L \to \infty$, all terms with $L$ on the denominators are neglectable:

$$\begin{aligned} \frac{1}{c_m}\frac{dc_m}{d\zeta} + \left\langle m|m\right\rangle \\ = \sum_{n\neq m} -\frac{\beta_n/\beta_m}{2\left(\beta_m^2-\beta_n^2\right)}\frac{c_n}{c_m} e^{i(\theta_m-\theta_n)}\left\langle m\right|\frac{dH}{d\zeta}\left|n\right\rangle\end{aligned} \tag{19}$$

In which we use (12). Integrate (19), we get:

$$\begin{aligned} \ln c_m\Big|_0^\zeta + \int_0^\zeta d\zeta_1 \left\langle m\right|\frac{d}{d\zeta_1}\left|m\right\rangle \\ = -\sum_{n\neq m}\int_0^\zeta d\zeta_1 \frac{\beta_n/\beta_m}{2\left(\beta_m^2-\beta_n^2\right)}\frac{c_n}{c_m} e^{i(\theta_m-\theta_n)}\left\langle m\right|\frac{dH}{d\zeta_1}\left|n\right\rangle\end{aligned} \tag{20}$$

Then we take the second approximation. For any $f(\zeta)$:

$$\int_0^{\zeta} d\zeta_1 f(\zeta_1) e^{i(\theta_m-\theta_n)} = \frac{1}{iL}\int_0^{\zeta} \frac{f(\zeta_1)}{\beta_m-\beta_n} de^{i(\theta_m-\theta_n)}$$

$$= \left.\frac{f(\zeta_1)e^{i(\theta_m-\theta_n)}}{iL(\beta_m-\beta_n)}\right|_0^{\zeta} + \frac{1}{iL}\int_0^{\zeta} e^{i(\theta_m-\theta_n)} d\frac{f(\zeta_1)}{(\beta_m-\beta_n)}$$

$$= o(1/L) + \frac{1}{iL}\int_0^{\zeta} d\zeta_1 e^{i(\theta_m-\theta_n)} \tag{21}$$

$$\frac{f'(\zeta_1)(\beta_m-\beta_n) - f(\zeta_1)(\beta_m'-\beta_n')}{(\beta_m-\beta_n)^2}$$

$$= o(1/L) + o(1/L^2) + \ldots$$

So, RHS in (20) is first-order infinitesimal of $L$, so the final expression of $c_m$ becomes:

$$\ln c_m\big|_0^{\zeta} = i\int_0^{\zeta} d\zeta_1 i\langle m|\frac{d}{d\zeta_1}|m\rangle \tag{22}$$

$$c_m(\zeta) = c_m(0)e^{i\int_0^{\zeta} d\zeta_1 i\langle m|\frac{d}{d\zeta_1}|m\rangle} = c_m(0)e^{i\gamma(\zeta)} \tag{23}$$

Thus, we get a conclusion the same as AQA with $\gamma(\zeta)$ being real according to (13) and corresponding to the Berry Phase [39]. Obviously, $|c_m|^2$ is a constant, which means that the power won't exchange between eigenmodes.

Consider the terms that have been eliminated in the second approximation and integrate them by parts:

$$\int_0^{\zeta} d\zeta_1 \frac{\beta_n/\beta_m}{2(\beta_m^2-\beta_n^2)} \frac{c_n}{c_m} e^{i(\theta_m-\theta_n)} \langle m|\frac{dH}{d\zeta_1}|n\rangle$$

$$= \int_0^{\zeta} \frac{\beta_n/\beta_m}{iL(\beta_m^2-\beta_n^2)(\beta_m-\beta_n)} \frac{c_n}{c_m} \langle m|\frac{dH}{d\zeta_1}|n\rangle de^{i(\theta_m-\theta_n)} \tag{24}$$

$$= \left.\frac{e^{i(\theta_m-\theta_n)}\beta_n/\beta_m}{iL(\beta_m^2-\beta_n^2)(\beta_m-\beta_n)} \frac{c_n}{c_m} \langle m|\frac{dH}{d\zeta_1}|n\rangle\right|_0^{\zeta} + o(1/L^2)$$

So, the adiabaticity condition could be written as:

$$\begin{aligned} L &\gg \left|\frac{e^{i(\theta_m-\theta_n)}\beta_n/\beta_m}{(\beta_m^2-\beta_n^2)(\beta_m-\beta_n)} \langle m|\frac{dH}{d\zeta}|n\rangle\right| \\ &= \frac{\beta_n/\beta_m}{(\beta_m^2-\beta_n^2)(\beta_m-\beta_n)} \left|\langle m|\frac{dH}{d\zeta}|n\rangle\right| \end{aligned} \tag{25}$$

Where $c_n/c_m$ is not included in the condition as its module is constant and does not impact our algorithm. Substitute $|m\rangle, |n\rangle$ and $\widehat{H}$ in the (25):

$$\text{RHS} = \frac{\beta_n/\beta_m}{\left(\beta_m^2-\beta_n^2\right)\left(\beta_m-\beta_n\right)} \left|\iint_S d\mathbf{s}\left(\mathbf{E}_m \times \frac{d\mathbf{k}^2}{d\zeta}\mathbf{H}_n^* + \frac{d\mathbf{k}^2}{d\zeta}\mathbf{E}_n^* \times \mathbf{H}_m\right)\right|$$

$$= \frac{\lambda}{2\pi}\frac{n_{\text{eff},n}/n_{\text{eff},m}}{\left(n_{\text{eff},m}+n_{\text{eff},n}\right)\left(n_{\text{eff},m}-n_{\text{eff},n}\right)^2} \left|\iint_S d\mathbf{s}\left(\mathbf{E}_m \times \frac{d\mathbf{k}^2}{d\zeta}\mathbf{H}_n^* + \frac{d\mathbf{k}^2}{d\zeta}\mathbf{E}_n^* \times \mathbf{H}_m\right)\right| \tag{26}$$

Using $\mathbf{k}^2 = \omega^2\boldsymbol{\varepsilon}\boldsymbol{\mu} = (2\pi/\lambda)^2\boldsymbol{\varepsilon}_r$, we get the adiabaticity condition as:

$$L \gg C\frac{\left|\iint_S d\mathbf{s}\left(\mathbf{E}_m \times \frac{d\boldsymbol{\varepsilon}_r}{d\zeta}\mathbf{H}_n^* + \frac{d\boldsymbol{\varepsilon}_r}{d\zeta}\mathbf{E}_n^* \times \mathbf{H}_m\right)\right|}{\Delta n_{\text{eff},mn}^2} \tag{27}$$

Where:

$$C = \frac{\lambda}{2\pi}\frac{n_{\text{eff},n}/n_{\text{eff},m}}{n_{\text{eff},m}+n_{\text{eff},n}}, \Delta n_{\text{eff},mn} = n_{\text{eff},m} - n_{\text{eff},n} \tag{28}$$

$C$ usually changes very slowly, so the parts of the integration and the $\Delta n_{\text{eff},mn}$ matter more in the design of STA. Next, we will demonstrate the design strategy with the example of the PSR.

After defining the basic structure (see Section III in the main text), the integration in (27) could be further simplified:

$$\text{Int} = \left|\iint_S d\mathbf{s}\left(\mathbf{E}_m \times \frac{d\boldsymbol{\varepsilon}_r}{dw_1}\mathbf{H}_n^* + \frac{d\boldsymbol{\varepsilon}_r}{dw_1}\mathbf{E}_n^* \times \mathbf{H}_m\right)\right|\frac{dw_1}{d\zeta} \tag{29}$$

Where $d\boldsymbol{\varepsilon}_r/dw_1$ can be expressed as:

$$\frac{d\boldsymbol{\varepsilon}_r}{dw_1} = \Delta\boldsymbol{\varepsilon}_r\frac{\delta\left[(x,y)\in(l\times dw_1)\right]}{dw_1} \tag{30}$$

Where $\Delta\boldsymbol{\varepsilon}_r = \boldsymbol{\varepsilon}_r(\text{core}) - \boldsymbol{\varepsilon}_r(\text{clad})$, and $(l \times dw_1)$ represents the quadrangular area defined by the integration line and $dw_1$ shown in Fig. 1(c) and $\delta$ is the Dirac function. Equation (30) indicates an abrupt change of permittivity in the quadrangle, and the fact that $dw_1$ is a differentiation symbol degenerates the surface integration into a line integration as:

$$\text{Int} = \sin\theta\left|\int_l\left(\mathbf{E}_m \times \Delta\boldsymbol{\varepsilon}_\mathbf{r}\mathbf{H}_n^* + \Delta\boldsymbol{\varepsilon}_\mathbf{r}\mathbf{E}_n^* \times \mathbf{H}_m\right)\mathbf{z}ds\right|\frac{dw_1}{d\zeta} \tag{31}$$

Where $\theta$ is the sidewall angle of the waveguides and $\mathbf{z}$ is the unit vector of propagation direction. Define:

$$
\begin{aligned}
A_{mn} &\equiv \frac{\left|\int_L \left(\mathbf{E}_m \times \Delta\boldsymbol{\varepsilon}_{\mathbf{r}} \mathbf{H}_n^* + \Delta\boldsymbol{\varepsilon}_{\mathbf{r}} \mathbf{E}_n^* \times \mathbf{H}_m\right) \mathbf{z} dl\right|}{\Delta n_{\mathrm{eff},mn}^2} \frac{dw_1}{d\zeta} \\
&= \frac{I_{mn}}{\Delta n_{\mathrm{eff},mn}^2} \frac{dw_1}{d\zeta}
\end{aligned} \tag{32}
$$

Putting $\sin\theta$ into $C$, the adiabaticity condition could be concisely expressed as:

$$
L \gg C \cdot A_{mn} \tag{33}
$$